\PassOptionsToPackage{table}{xcolor}

\documentclass[10pt,compsoc, journal]{IEEEtran}

\usepackage{dblfloatfix}
\usepackage[utf8]{inputenc}
\usepackage[breakable]{tcolorbox}

\usepackage{rotating}
\usepackage{booktabs}
\usepackage{wrapfig}
\usepackage{multirow}
\usepackage{balance}
\usepackage[shortlabels]{enumitem}
\usepackage{ragged2e}
\usepackage{tabularx}
\usepackage{svg}
\usepackage{pifont}
\usepackage{graphicx}
\usepackage[numbers,sort]{natbib}
\usepackage{colortbl}
\usepackage{multirow}
\usepackage{adjustbox}
\usepackage{picture}
\usepackage{subcaption}
\usepackage[ruled,linesnumbered]{algorithm2e}
\usepackage{url}
\usepackage{amsmath}
\usepackage{amsthm}
\usepackage{amsfonts}
\usepackage[table]{xcolor}
\newtcolorbox{blockquote}{colback=blue!5!white,boxrule=0.4pt,colframe=gray!60!black,fonttitle=\bfseries}
\newcolumntype{L}{>{\raggedright\arraybackslash}X}
\newcommand{\bi}{\begin{itemize}}
\newcommand{\ei}{\end{itemize}}

\newcommand{\tick}{\ding{51}}

\newcommand{\quart}[3]{\begin{adjustbox}{max width=.2\linewidth}\begin{picture}(100,5)%1
    {\color{black}\put(#3,2){\circle*{7}}\put(#1,2){\line(1,0){#2}}}\end{picture}\end{adjustbox}}

\usepackage[tikz]{bclogo}
{\noindent\begin{minipage}[c]{\linewidth}%
\begin{bclogo}[couleur=gray!20,%
                arrondi=0,logo=\none,% 
                ombre=true%
                ]{{\small  ~#1}}}%
{\end{bclogo}\vspace{2mm}\end{minipage}}

\newcommand{\BLUE}{\color{black}}
\newcommand{\BLACK}{\color{black}}

\title{
On the Value of Oversampling for Deep Learning in Software Defect Prediction
}
\author{Rahul~Yedida and Tim~Menzies,~\IEEEmembership{IEEE Fellow}%
\IEEEcompsocitemizethanks{\IEEEcompsocthanksitem R. Yedida and T. Menzies are with the Department of Computer Science, North Carolina State University. Email: ryedida@ncsu.edu, timm@ieee.org}.%
\thanks{Manuscript submitted to TSE,  August 9, 2020.}}
\begin{document}

 \definecolor{blueblue}{RGB}{16,52,109}
\definecolor{redred}{RGB}{101,0,26}

\IEEEtitleabstractindextext{
\begin{abstract}
One truism of deep learning is that the automatic feature engineering (seen in the first   layers of those
networks) excuses  data  scientists  from performing   tedious  manual  feature engineering prior to running DL.
For the specific case of deep learning for defect prediction, we show that that truism is false. 
Specifically,
when we pre-process data with a novel oversampling technique called fuzzy sampling, as part of a larger pipeline called GHOST (Goal-oriented Hyper-parameter Optimization for Scalable Training),  then we can do significantly
better 
than the prior DL state of the art in 14/20 defect data sets. 
Our approach yields state-of-the-art results significantly faster deep learners. These results present a cogent case for the use of oversampling prior to applying deep learning on software defect prediction datasets.
\end{abstract}
}

\maketitle
%\IEEEdisplaynontitleabstractindextext

\section{Introduction}
\label{sec:intro}

Can  deep learning (DL) be applied to SE data without first
adjusting those learners to the particulars of SE?
Many researchers believe so.
A common claim is that DL supports a kind of 
automated feature engineering~\cite{zeiler2014visualizing,panda2016unsupervised,nair2010rectified,suk2014hierarchical,yamashita2018convolutional} 
that lets  data scientists   avoid
tedious manual feature engineering,  prior to running their learners.
\begin{figure*}
\begin{tabular}{cc}
Figure~\ref{b4after}.a {\bf F1 scores before} (off-the-shelf deep learning) & 
Figure~\ref{b4after}.b
{\bf F1 scores after}  (using the methods of this paper)\\ 
\includegraphics[width=2.8in]{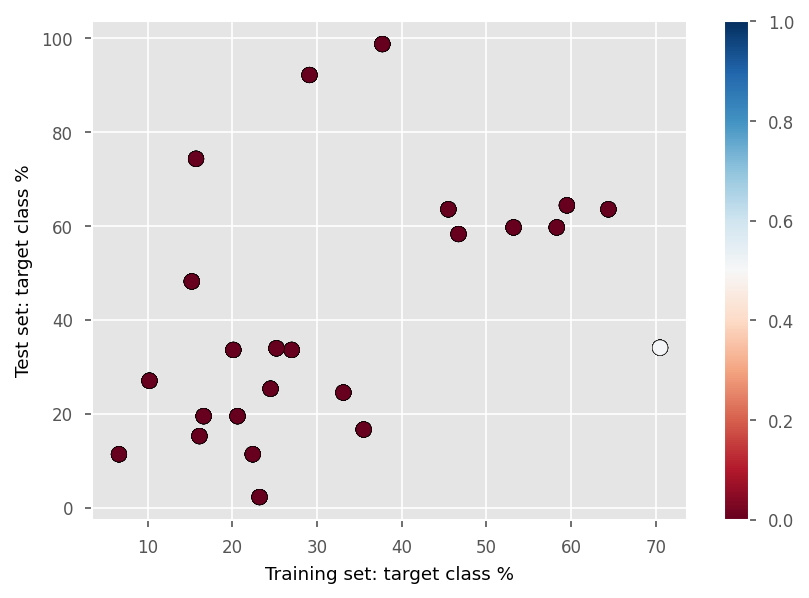} &
\includegraphics[width=2.8in]{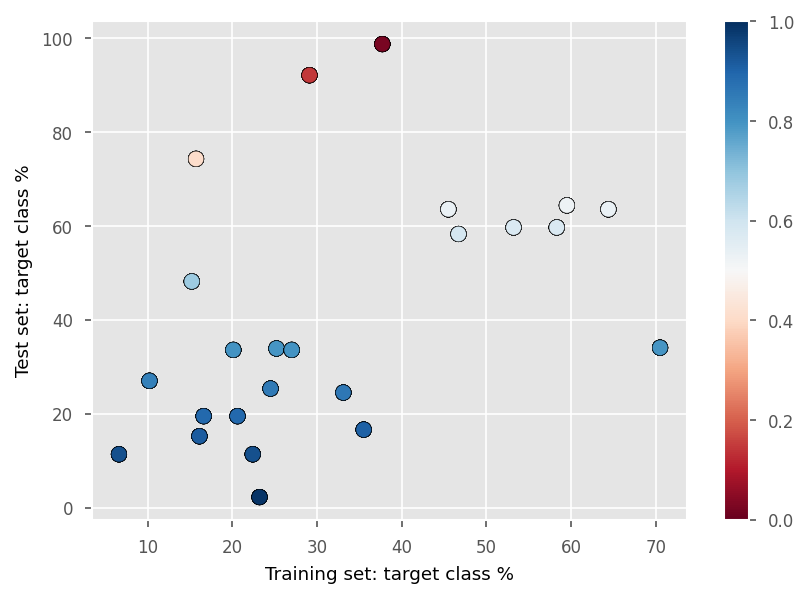}
\end{tabular}
\caption{Defect prediction  using project data described
later in this paper.  Data  is divided 70:30 into train:test sets.
Performance scores come from applying a model
(learned from  training data) to the   test data. In this
figure 
\colorbox{blueblue}{\textcolor{white}{{\bf blue}}}
dots are good
\colorbox{redred}{\textcolor{white}{{\bf red}}}
 are bad. 
 Figure~\ref{b4after}.a and
  Figure~\ref{b4after}.b were
  created using the technology
  described in lines \#1 and \#8 (respectively) of 
 Table~\ref{tab:ghost-stages} (described. below). To allow  comparison between   figures,  Figure~\ref{b4after}.b shows
 (i)~ the imbalance {\em before} applying the methods of this paper
 and (ii)~the  Figure~\ref{b4after}.b  performance scores  come after 
 {\em after} we applied the methods of this paper. }\label{b4after}
\end{figure*}

It is timely to assess such claims.
DL is now
widely applied to  many SE tasks such as   
bug localization~\cite{huo2019deep}, 
sentiment analysis~\cite{lin2018sentiment,guo2017semantically}, 
API mining~\cite{chen2019mining,gu2016deep,nguyen2017exploring}, effort estimation for agile development~\cite{choetkiertikul2018deep}, code similarity detection~\cite{zhao2018deepsim},
code clone detection~\cite{white2016deep}, etc.
%Some researchers are worried that DL is being applied with insufficient critical assessment. 
% For example, Gary Marcus~\cite{marcus2018deep} 
% warns than DL may be so over-hyped that it runs  ``fresh risk for seriously dashed expectations'' that could blinker AI researchers from trying new ideas as well as
% bringing  another AI winter (the period in the mid to late 1980s when most AI companies went bankrupt due to poor generalization
% of their methods).
%  A major issue with deep learning is it high  computationally expensive. For example, in the taking CMA-ES \cite{loshchilov2016cma} for example, they take 17 hours to run hyperparameter optimization on 30 GPUs (running each deep learner for 30 minutes). Such resources are not available to all research groups; we seek to improve this for software engineering. 
%So what is the benefit of DL? Is the technology over-hyped? 
%Should we stop using everything else and just move to DL?
Despite its widespread use, few SE researchers  critically examining the utility of deep learning for SE.
In their literature review, Li et al.~\cite{li2018deep} explored over 40 SE papers facilitated by DL models and found that 33 of them used  DL without comparison to   other (non-DL) techniques. 
In our own literature review on DL in SE (reported below in \S\ref{sec:background}), we       find    experimental
comparisons of DL-vs-non-DL in  less than 10\% of    papers.

This paper shows that a non-DL technique called  ``oversampling''
(artificially generating members of a minority class {\em prior} to running a learner)  dramatically
improves  deep learning.
Such resampling is {\em not} widespread practice.
For example, Wang et al. \cite{wang2016automatically}
recently achieved state-of-the-art performance in defect prediction. Their paper did not discuss or deploy any  class imbalance techniques--a pattern that repeats across all
the papers seen in our literature review.
This is a limitation with current research since:
\bi
\item
When we tried their methods using static
code attributes we found that
(a)~many data sets were imbalanced (target classes under 30\% or
even less) and (b)~standard DL performed quite poorly (see  all the \colorbox{redred}{\textcolor{white}{{\bf red}}} dots in  Figure~\ref{b4after}.a).
\item
But applying our sampling methods improved performance  dramatically 
 (see  the \colorbox{blueblue}{\textcolor{white}{{\bf blue}}} dots of Figure~\ref{b4after}.b).
 \ei

 The rest of this paper discusses deep learning and  novel oversampling methods that can improve its performance.
  Our case study is defect prediction from \textit{static code features} (e.g. depth of inheritance trees, class coupling and cohesion, lines of code, etc) to  guide human inspection effort towards small regions
 of the code base that are most likely to have errors. Our argument will proceed as follows: we will first build a case for the problems with learning from a class-imbalanced dataset. Then, we will discuss approaches used in the literature for alleviating these problems. In particular, we will discuss the approach studied by Buda et al. \cite{buda2018systematic}. We will then show that for defect prediction, this approach is insufficient, and extend their method to a novel \textit{adaptive} oversampling approach. We will show that when used as part of a larger framework called GHOST, we achieve state-of-the-art results significantly faster. Our ablation study will demonstrate that the biggest leaps in performance come from our novel oversampling approach, which is the core component of GHOST.
 
Before starting, we digress to make three points:
 \bi
 \item It is important 
 not to overstate our results.
 This paper only shows that oversampling improves deep learning for defect prediction. That said, the results of this paper motivate
 a new approach to deep learning where researchers check if those algorithms can be extended and improved by  support tools like our adaptive oversampling.
 \item
Any paper  recommending oversampling
 needs to  document that it  avoids the following  threat to validity. While we oversample the training data, we 
 {\em never} modify the test data (since that would we mean are not testing on the kinds of data that might be encountered  in the field).
 \item
 To support open science, all the scripts and data used in this study are freely available online\footnote{\url{https://tiny.cc/ghost-dl}}.
\ei

\section{Background}
\label{sec:background}
\subsection{Defect Prediction}
Every quality assurance decision is associated with a human and resource cost to the developer team.
It is impractical and inefficient to distribute equal effort to every component in a software system~\cite{briand1993developing}.
Hence, 
it is common to match the quality assurance (QA) effort to the perceived criticality and bugginess of the code for managing resources efficiently. 
 
  {\em Defect prediction models}~\cite{fu2016tuning} are one way to  take a look at the incoming changes and focus inspection effort on specific modules (ignoring others).
  There is much  industrial interest in such predictors:
 a  survey  of  395  practitioners  from  33  countries  and five  continents~\cite{wan2018perceptions} found  that  over  90\%  of the  respondents  were  willing  to  adopt  defect  prediction techniques.
Such predictors can save labor compared with traditional manual methods.
For a telecommunications company, 
Misirli et al.~\cite{misirli2011ai} built a defect prediction model that predicted 87\% of code defects and decreased inspection efforts by 72\% (and reduced post-release defects by 44\%). 
Kim et al.~\cite{kim2015remi} applied  a
defect prediction model to   API development process at Samsung.
Their models could
predict the bug-prone APIs with reasonable accuracy~(0.68 F1 scores) and reduce the resources required for executing test cases. 
These defect predictors  require very simple code features
and hence are very cheap to build (compared to using, say, a static code analysis tool with an expensive license)
And while these models use simple attributes, they are still remarkably effective.
Rahman et al. ~\cite{rahman2014comparing} compared (a) static code analysis tools FindBugs, Jlint, and PMD with (b) defect predictors (which they called ``statistical defect prediction'') built using logistic regression.
No significant differences in cost-effectiveness were observed.
\subsection{Oversampling}\label{over}

Building classification models for, say, defect prediction  
becomes complicated~\cite{agrawal2018better,4288197,Chawla02} when   the target class is rare (since it is hard for the learner to build a model that can find that target).
To solve that problem,
much prior work as argued for {\em oversampling}, augmented with some {\em tuning}, as a method for improving learner performance.

Buda et al.~\cite{buda2018systematic} note that in many cases, simple oversampling suffices; i.e.,
many times,
pick one example at random
then add that it several
times
back into the training set.
Other researchers take a different view.
Chawla~\cite{Chawla02} showed that by generating synthetic examples of a scare minority class,
it is possible to boost learner performance. In their SMOTE algorithm, examples in the minority class build synthetic examples
halfway between themselves and their near neighbors.

Oversampling alleviates  class imbalance  by drawing additional samples from the minority class. As shown in Figure \ref{b4after}, in datasets with a class imbalance, the performance of the learner drops drastically; only in a case with less severe imbalance (training set imbalance = 70\%) does the learner achieve moderate performance. A standard method of doing this is by simply duplicating the points in the minority class until the number of samples in both classes is the same. Oversampling is not the only way to fix the class imbalance problem, however. Another common approach is undersampling, where the majority class is \textit{downsampled} to match the number of samples in the minority class.

% % \begin{figure}[hb]
% %     \centering
    
% %     \begin{subfigure}{0.49\linewidth}
% %         \includegraphics[width=\linewidth]{vanilla-results.png}
% %         \caption{}
% %     \end{subfigure}
% %     \begin{subfigure}{0.49\linewidth}
% %         \includegraphics[width=\linewidth]{final-results.png}
% %         \caption{}
% %     \end{subfigure}
% %     \caption{Performance (measured by F1-score) of a neural network on our defect prediction datasets. \textbf{(a) Before:} Results with a ``vanilla" network, i.e., with none of the techniques from \S \ref{sec:ghost}. Note that with large class imbalance ratios, the performance is poor. \textbf{(b) After:} Results after applying the techniques from our method.}
% %     \label{fig:vanilla-results}
% % \end{figure}

%Although undersampling yields fewer total samples, which can speed up the runtimes of algorithms, we found that in several of our datasets, the minority class had too few samples ($<50$), which would have reduced our datasets to too few samples, and worse still, this would lead to test data whose distributions may not be representative of the population (since we do not perform oversampling on test data). For this reason, we chose to use oversampling. 

\subsection{Oversampling and Tuning}\label{tuneit}

Oversampling algorithms like SMOTE come with various control parameters that have to
be set via ``engineering judgement'' (i.e. guessing). Researchers like  Agrawal et al.~\cite{agrawal2018better} augment SMOTE with tuning algorithms that automatically
find those  parameters. There are many such ``hyperparameter optimziation'' methods such as MOEA/D, NSGA-II, differential evolution, hyperopt, etc~\cite{agrawal2019dodge,bergstra2013making,bergstra2011algorithms}.
In subsequent work, Agrawal et al.~\cite{agrawal2020simpler} found a measure
(that they call ``intrinsic dimensionality'') that can be used to select  which hyperparameter optimizer
 works best for
different kinds of data. 
For example, based on their analysis, this paper
does not use Hyperopt~\cite{bergstra2011algorithms} (since that algorithm tends to
overfit on our kind of data). Instead, following 
Agrawal et al.'s advice,
we will use the   DODGE~\cite{agrawal2019dodge}  hyperparameter optimizer to tune our oversamplers. 
DODGE experiments with  a set of  tunings $T_i$ in a sequential ordering $T_1,T_2 \ldots$ etc. If any tuning $T_i$ results in a performance near some prior tuning $T_{j<i}$ then DODGE marks that tuning as ``tabu'' and  avoids  any similar tuning.
For two reasons, DODGE runs very quickly.
Firstly the tabu space grows,
this kind of search very quickly runs out of new things to try
(hence DODGE can terminate after just a few dozen experiments).
Secondly, prior to doing any tuning, Agrawal et al. reduce
the feature space using  {\em feature selection}. Feature selectors
prune superfluous attributes (i.e. those not associated with the the target class).
Following the advice of Hall and Holmes~\cite{holmes03}, DODGE
uses Hall's   CFS    selector~\cite{hall00}\footnote{
CFS favors   feature subsets that are weakly associated with 
each other and strongly associated with the target class.
This is computed using 
$
\mathit{merit}_s = \frac{k\overline{r_{\mathit{cf}}}}{\sqrt{k+k(k-1)\overline{r_{\mathit{ff}}}}}
$
where:
$\mathit{merit}_s$ is the value of some subset $s$ of the
features containing $k$ features; 
$r_{\mathit{cf}}$ is a score describing the connection of that feature
set to the class;
and $r_{\mathit{ff}}$ is the mean score of the feature to feature
connection between the items in $s$.
Note that for this to be maximal, $r_{\mathit{cf}}$  must be large
and $r_{\mathit{ff}}$ must be small. That is, features have to connect
more to the class than each other. 
}. Finally, we found that using SMOTE \cite{Chawla02} improved the results of DODGE, so we used it to provide a stronger baseline.

\subsection{Deep Learning}
\label{sec:deeplearning}
Deep learning  (DL) is an extension of prior work
on neural networks where the 
 adjective ``deep'' refers to the use of multiple layers in the network. 
In the 1960s and 1970s it was  found that 
very simple neural nets  can be poor classifiers
unless they are extended with
(a)~extra layers between inputs and outputs;
and (b)~a nonlinear activation function controlling links from inputs to a hidden
layer (which can be very wide) to an output layer.
Deep learning is a modern variation on the above which is concerned with a potentially unbounded number of layers of bounded size\footnote{There has been some theoretical interest in layers with unbounded sizes (see \cite{jacot2018neural}); however, this has not been used in practice.}. 
Last century,  most neural networks used the ``sigmoid" activation function ($f(x) = \frac{1}{1+e^{-x}}$), which was subpar to other learners in several tasks; it was only when the ReLU (rectified linear unit) activation function $f(x) = \max(0, x)$ was introduced by \cite{nair2010rectified} that their performance increased dramatically, and they rose to popularity. To train a deep learner:
\bi
\item
Forward-propagation
finds outputs from the  inputs.
\item If the weights learned at layer $i$ (which together form the parameters for the model) are represented as $W^{[i]}$ and $b^{[i]}$ respectively, then

\begin{align*}
    a^{[l]} &= f(W^{[l]T}a^{[l-1]}+b^{[l]}) \\
    a^{[0]} &= X \\
    a^{[L]} &= y
\end{align*}

where $L$ is the total number of layers. 
\item
Next, a ``backpropagation" \cite{rumelhart1986learning} approach, first defined in 1986, is used to learn the parameters.
\item
This process is repeated for a certain number of ``epochs", which is typically chosen \textit{a priori}.
\ei
This paper will compare our new oversampling methods against two DL baselines: a {\em standard deep learner} (derived from a literature review)  and a {\em state-of-the-art}  DL 
defect predictor that we take from the recent SE literature.

For the  {\em standard deep learner}, we use the kind
 of learners whose  architecture would be 
 ``reasonably justifiable''  to a deep learning expert. 
 To determine ``reasonably justifiable'', we used the results of a recent paper published in a top venue (NeurIPS) by Montufar et al. \cite{montufar2014number}, who derive theoretical lower and upper bounds on the number of piecewise linear regions of the decision boundary learned by a deep learner.
By maintaining the number of nodes in the hidden layers of the network to be at least as many as the number of inputs (i.e., the dimensionality of the data),
Montufar et al. show that the lower bound for the number of piecewise linear regions in the decision boundary is nonzero. We interpret this as, by setting the number of nodes in each hidden layer \textit{equal} (for simplicity) to the dimensionality of the data, the network must make an effort to separate the two classes (i.e., using a nonzero number of piecewise linear regions). This certainly does not guarantee an optimal boundary, but it does provide a guarantee for a nontrivial decision boundary.

 For the
 {\em state-of-the-art  DL}, we use the Wang et al. approach~\cite{wang2018deep,wang2016automatically}.
 This  is a Deep Belief Network (DBN) \cite{hinton2009deep} that learns ``semantic" features and then uses classical learners to perform defect prediction.
 In that approach,
for each file in the source code, they extract tokens, disregarding ones that do not affect the semantics of the code, such as variable names. 
These tokens are vectorized and given unique numbers, forming a vector of integers for each source file. 
%These vector-buggy pairs form the training set, which are input to a Deep Belief Network (DBN), a neural network architecture designed to \textit{automatically} extract features. 
The features learned in this way  are used as input to a classical learner, such as Naive Bayes.  Note that, like much of the DL literature, Wang et al. assume
that their algorithm automates feature selection (so they do not make use
of any other feature engineering pre-processor).

\begin{figure}[!b]%{r}{2.1in}
     \begin{center}
    \includegraphics[width=.8\linewidth]{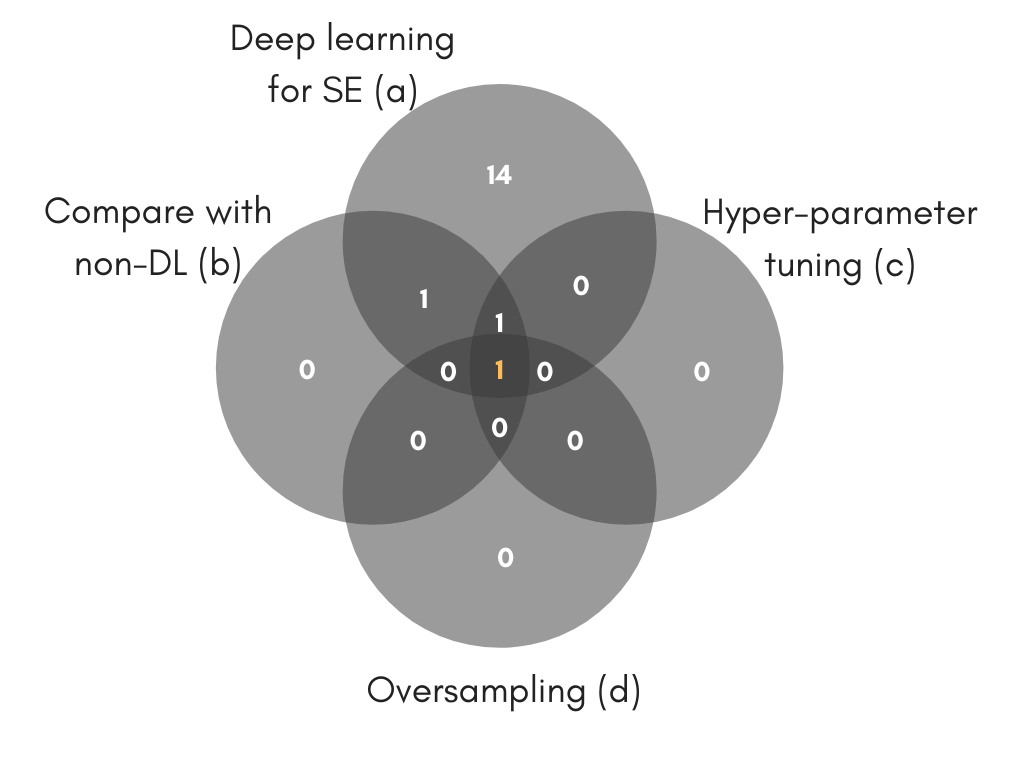}
       \end{center} 
    \caption{SE+DL  literature review.
    The paper satisfying all three criteria (i.e., $a + b + c$) is \cite{wang2016automatically}; the paper that compares with classical learners (i.e., $a + b$) is \cite{dam2019lessons}. 
    This paper is the only one to satisfy all four criteria, $a+b+c+d$ (denoted by the ``1'' in the center of this chart).}
    \label{fig:venn}
\end{figure}

\begin{table*}[t!]
\caption{Selected papers after applying filters (top SE venues, at least 10 cites per year).}\label{tab:papers}
\label{tab:selected-papers}
{\rowcolors{2}{white}{blue!5}
\scriptsize
\begin{tabularx}{\textwidth}{llrlL}
    \toprule
	Year & Venue & Cites & Use Cases & Title \\ 
	\midrule
	2016 & ICSE & 262 & defect prediction & Automatically Learning Semantic Features for Defect Prediction \cite{wang2016automatically} \\
	2016 & ASE & 224 & code clone detection & Deep learning code fragments for code clone detection \cite{white2016deep} \\
	2015 & MSR & 183 & code representations & Toward Deep Learning Software Repositories \cite{deshmukh2017towards} \\
	2017 & ICSE & 83 & trace links & Semantically Enhanced Software Traceability Using Deep Learning Technique \cite{guo2017semantically} \\
	2017 & ICSME & 58 & code clone detection & CCLearner: A Deep Learning-Based Clone Detection Approach \cite{li2017cclearner} \\
	2019 & TSE & 37 & story point estimation & A Deep Learning Model for Estimating Story Points \cite{choetkiertikul2018deep}\\ 
	2018 & MSR & 34 & code clone detection & Deep Learning Similarities from Different Representations of Source Code \cite{marcus2018deep} \\ 
	2017 & ICSME & 33 & vulnerability prediction & Learning to Predict Severity of Software Vulnerability Using Only Vulnerability Description \cite{han2017learning} \\
	2018 & TSE & 24 & defect prediction & Deep Semantic Feature Learning for Software Defect Prediction \cite{wang2018deep} \\ 
	2017 & ICSME & 23 & duplicate bug retrieval & Towards Accurate Duplicate Bug Retrieval Using Deep Learning Techniques \cite{deshmukh2017towards} \\ 
	2019 & ICSE & 10 & bug localization & CRADLE: Cross-Backend Validation to Detect and Localize Bugs in Deep Learning Libraries \cite{pham2019cradle} \\ 
	2019 & MSR & 8 & defect prediction & DeepJIT: An End-to-End Deep Learning Framework for Just-in-Time Defect Prediction \cite{hoang2019deepjit} \\
	2019 & MSR & 5 & defect prediction & Lessons Learned from Using a Deep Tree-Based Model for Software Defect Prediction in Practice \cite{dam2019lessons} \\ 
	2019 & ICSE-NIER & 4 & transfer learning & Leveraging Small Software Engineering Data Sets with Pre-Trained Neural Networks \cite{robbes2019leveraging} \\
	2018 & TSE & 4 & defect prediction & How Well Do Change Sequences Predict Defects? Sequence Learning from Software Changes \cite{wen2018well} \\ 
	2018 & IC SESS & 1 & language model & A Neural Language Model with a Modified Attention Mechanism for Software Code \cite{zhang2018neural} \\
 \bottomrule  
\end{tabularx}
}
\end{table*}

\subsection{Deep Learning (as seen in SE) }

To say the least, much of the technology advocated
by this paper is not standard practice
in  the DL literature.
Numerous prominent papers~\cite{zeiler2014visualizing,panda2016unsupervised,nair2010rectified,suk2014hierarchical,yamashita2018convolutional} and data science blogs\footnote{E.g. 2018: Introduction to Automated Feature Engineering Using Deep Feature Synthesis, \url{https://tinyurl.com/y3p9wgey}; Automated Feature Engineering, \url{https://tinyurl.com/y6ya8q5c}. 
2019: Why Automated Feature Engineering Will Change the Way You Do Machine Learning,   \url{https://tinyurl.com/y37ys5bd}.
2020: Automated Feature Engineering Tools, \url{https://tinyurl.com/y6qjyqnv}.}
say that DL offers ``automated feature engineering'' 
that removes the need for intricate preprocessing.
In that view, the first
few layers of a deep learning network automatically re-weight/ignore
important/irrelevant inputs~\cite{nair2010rectified}.

  Figure \ref{fig:venn} summarizes our recent
  review of DL in SE.
  Note that very few papers compare their results with non-deep learning methods \cite{wang2016automatically,dam2019lessons}  
 and none explored oversampling or the hyperparameter tuning methods discussed in \S\ref{tuneit}. To build that diagram, 
 using Google Scholar, we search for research papers using the keyword ``deep learning AND software engineering". This returned over 900 papers with at least one citation.  
To narrow down that list, we looked for papers   published in the top venues (defined as per   Google Scholar metrics "software systems"),  with $N\ge 10$ cites per year (and for papers
more recent than 2017  we used $N \ge 1$).  This led to 16 papers listed in Table~\ref{tab:selected-papers}.

Outside of the SE literature, we could find one highly cited paper arguing for  something like
our proposal.
Buda et al.~\cite{buda2018systematic} show
that it is very useful to augment
 DL with an oversampling preprocessor. 
According to Google Scholar, this paper has received 731 citations since its publication in 2017. This paper is discussed
in the next section.

(Aside: One  feature we note from this sample of  papers is that the code and data for most of these papers is not always open source. Some are even protected by patent applications. This has implications for reproducibility. For example, this paper
 baselines
 some of our
 results against
  Wang et al. \cite{wang2016automatically}.
  Their methods
  are protected by a patent, and therefore we could not replicate their results using their code.
  That said, we were able to 
  find their exact training and test sets,
  which we used
  for comparison purposes.)

\begin{table}[!t]
    \centering   
    \caption{Quartile charts of F1 scores with various treatments. Results are across 240 data points, i.e., 10 repeats, each over the 24 datasets (combined results of the DODGE data and the Wang et al. \cite{wang2016automatically} data). \textbf{Legend:} 
    \textbf{a} = off-shelf deep learning,
    \textbf{b} = oversampling as studied by Buda et al. \cite{buda2018systematic},
    \textbf{c} = SMOTE \cite{Chawla02}, 
    \textbf{d} = fuzzy sampling,
     \textbf{e} = hyper-parameter tuning.}\label{compare}
    \label{tab:ghost-stages}
    \rowcolors{2}{blue!10}{white}
\adjustbox{max width=.5\textwidth}{    \begin{tabular}{r|lllll|rr|c}
  &
    \begin{turn}{90}deep learning\end{turn} &
    \begin{turn}{90}oversampling\end{turn} &
   \begin{turn}{90}SMOTE\end{turn} & 
   \begin{turn}{90}fuzzy sampler\end{turn}&
     \begin{turn}{90}tuning\end{turn} & \textbf{Median} &
    \textbf{IQR} & 
   \textbullet  = median and  \\
   & a&b&c&d&e&(50th) & & lines shows IQR \\\hline
\#1    &   \tick & & & & & 0 & 11 & \quart{0}{0}{0} \\
 \#2    &   \tick & \tick & & & & 0 & 4 & \quart{0}{0}{0} \\
\#3 & \tick & \tick & \tick & \tick & & 0 & 15 & \quart{0}{8.5}{0} \\
\#4 & \tick & \tick & \tick & & \tick & 0 & 29 & \quart{0}{30.4}{0} \\
%   \#3    &  \tick & & & \tick &  & 0 & 9 & \quart{0}{12.1}{0} \\
%   \#4    & \tick & & & & \tick & 0 & 30 & \quart{0}{40.9}{0} \\
%     \#5    &\tick & \tick & & & \tick & 0 & 30 & \quart{0}{31.8}{0} \\
%  \#6   &    \tick & & \tick & & \tick & 0 & 30 & \quart{0}{29.3}{0} \\
 \#5 & \tick & & \tick & \tick & \tick & 51 & 25 & \quart{32.6}{42.2}{50.7} \\
\#6 &        \tick & \tick & & \tick & \tick & 51 & 25 & \quart{32.6}{42.9}{51.2} \\
\#7 &       \tick & \tick & \tick & \tick & \tick & 51 & 25 & \quart{32.6}{42.9}{50.7} \\ 
GHOST = \#8 &      \tick & \tick & \tick & \tick\tick & \tick & 80 & 25 & \quart{57.4}{31.8}{79.6} 
    \end{tabular}}
\end{table}

 \section{Designing our Algorithm (GHOST)}\label{sec:algdesign}
 \label{sec:ghost} 
In this section, we describe the components that make up our approach, which we call GHOST\footnote{\textbf{G}oal-oriented \textbf{H}yper-parameter \textbf{O}ptimization for \textbf{S}calable \textbf{T}raining.}.
  For a detailed statistical analysis of our results,
compared to two baselines, see the {\em next} section.

Table~\ref{compare} shows the relative effectiveness of the components within GHOST. Each row
is a different algorithm made up of the components listed in the columns.
Several of those components have already been described.
For example, {\em deep learning} is the ``standard deep learner'' from \S\ref{tuneit}.
Also, the
SMOTE and tuning components of  Table~\ref{compare} were discussed above in \S\ref{over} and \S\ref{tuneit}. 
Note that Figure~\ref{fig:wfo:none} and Figure~\ref{fig:wfo:final} came from line \#1 and \#8 of Table~\ref{compare}.

The other components are described in this section.
{\em Oversampling} is a technique taken from  Buda et al.~\cite{buda2018systematic}.
Also,  the  {\em fuzzy sampler}  is a novel method developed for this paper.
Note that row~\#8 has two ticks \tick\tick  in column d; i.e. in our preferred
approach, the fuzzy sampler can be  applied twice (see the discussion on line 32 of Algorithm~\ref{alg:ghost}, below, for details).

The last two columns  of Table~\ref{compare} show the  median F1 (the harmonic mean of precision and recall) seen across all the data of Table~\ref{tab:datasets}.
Note that, in Table~\ref{compare}, ``IQR'' denotes the (75-25)th interval of F1 scores.
In this table, we observe that:
\bi
\item 
The first row
shows the results presented in Figure~\ref{b4after}.a;
\item
 The last row shows the results of Figure~\ref{b4after}.b;
 \item
All the other rows show {\em ablation studies} where we took out one component from row~\#8. As can be seen, those ablation studies demonstrated that all the components of row~\#8 are required
to reach our best performance. 
\ei
The rest of this section discusses the component technologies used in  Table~\ref{compare}.

\begin{table}[!t]
    \centering
    \caption{ Software projects data used in this paper \cite{agrawal2019dodge}.
    For an explanation of attributes used in this data, see Table~\ref{tab:attributes}.}
    { \scriptsize
    \begin{tabularx}{\linewidth}{lllLL}
        \toprule
        Project & Train versions & Test versions & Training Buggy \% & Test  Buggy \% \\
        \midrule
        ivy & 1.1, 1.4 & 2.0 & 22 & 11 \\
        \rowcolor{blue!10} lucene & 2.0, 2.2 & 2.4 & 53 & 60 \\
        poi & 1.5, 2.0, 2.5 & 3.0 & 46 & 65 \\
        \rowcolor{blue!10}synapse & 1.0, 1.1 & 1.2 & 20 & 34 \\
        velocity & 1.4, 1.5 & 1.6 & 71 & 34 \\
        \rowcolor{blue!10}camel & 1.0, 1.2, 1.4 & 1.6 & 21 & 19 \\
        jEdit & 3.2, 4,0, 4.1, 4.2 & 4.3 & 23 & 2 \\
        \rowcolor{blue!10}log4j & 1.0, 1.1 & 1.2 & 29 & 92 \\
        xalan & 2.4, 2.5, 2.6 & 2.7 & 38 & 99 \\
        \rowcolor{blue!10}xerces & 1.2, 1.3 & 1.4 & 16 & 74 \\
        \bottomrule \\
    \end{tabularx} }
    \label{tab:datasets}
\end{table}
\begin{table}[!t]
    \centering
    \caption{Static attributes for the classes
    seen in the Table~\ref{tab:datasets} data.}
    { \scriptsize
    \begin{tabularx}{\linewidth}{lL}
    \toprule
        Attribute   & Description (with respect to a class) \\
        \midrule
        wmc & Weighted methods per class \cite{chidamber1994metrics} \\
    \rowcolor{blue!10}    dit & Depth of inheritance tree \cite{chidamber1994metrics} \\
        noc & Number of children \cite{chidamber1994metrics} \\
    \rowcolor{blue!10}    cbo & Coupling between objects \cite{chidamber1994metrics} \\
        rfc & Response for a class \cite{chidamber1994metrics} \\
    \rowcolor{blue!10}    lcom & Lack of cohesion in methods \cite{chidamber1994metrics} \\
        lcom3 & Another lcom metric proposed by Henderson-Sellers \cite{henderson1996coupling} \\
    \rowcolor{blue!10}    npm & Number of public methods \cite{bansiya2002hierarchical} \\
        loc & Number of lines of code \cite{bansiya2002hierarchical} \\
    \rowcolor{blue!10}    dam & Fraction of private or protected attributes \cite{bansiya2002hierarchical} \\
        moa & Number of fields that are user-defined types \cite{bansiya2002hierarchical} \\
    \rowcolor{blue!10} mfa & Fraction of accessible methods that are inherited \cite{bansiya2002hierarchical} \\
        cam & Cohesion among methods of a class based on parameter list \cite{bansiya2002hierarchical} \\
    \rowcolor{blue!10} ic & Inheritance coupling \cite{tang1999empirical} \\
        cbm & Coupling between methods \cite{tang1999empirical} \\
    \rowcolor{blue!10} amc & Average method complexity \cite{tang1999empirical} \\
        ca & Number of classes depending on a class \cite{tang1999empirical} \\
    \rowcolor{blue!10} ce & Number of classes a class depends on \cite{tang1999empirical} \\
        max\_cc & Maximum McCabe's cyclomatic complexity score of methods \cite{mccabe1976complexity} \\
    \rowcolor{blue!10} avg\_cc & Mean of McCabe's cyclomatic complexity score of methods \cite{mccabe1976complexity} 
    \end{tabularx} }
    \label{tab:attributes}
\end{table}

Row \#1 of Table~\ref{compare} shows results from applying the standard DL (of \S\ref{sec:deeplearning}) to defect prediction.  Those results are very low (median of zero), an effect we can explain as follows.

As seen in  Table~\ref{tab:datasets}, there is wide variability in the percent of defects seen in the training and test sets. For example, the defect ratios in velocity's  train:test ratios decrease from 71 to 34\%; jEdit's train:test sets ratios decrease from  23 to 2\%; xerces' train:test sets ratios increase from  16 to 74\%.
Such massive changes in the target class ratios means
that the geometry of the hyperspace boundary
between different classifications  
can alter dramatically between train and test.
Therefore, the ``appropriate
learner''   for
this kind of data is one that
works well for a wide range
of ratios
of class distributions.

More importantly, we note that in several of these datasets, there is a significant class imbalance problem. From the previous section, our first instinct was to use Buda et al.'s oversampling method to rectify this issue. 

Row \#2 of Table~\ref{compare} shows results from applying the oversampling methods
  of Buda et al. Specifically, we apply the random oversampling techniques studied by Buda et al. to make the number of samples in the minority class equal to the majority class. In our experiments, we mimic the random oversampling studied by Buda et al. using a common technique, weighted loss functions (see \S \ref{sec:loss}). This approach, while having the same effect as random oversampling, is significantly computationally cheaper, and therefore, preferred. 

As seen in  row \#2 of Table~\ref{compare}, the results from applying the
Buda et al. methods are actually worse than Row \#1. 
To explain this negative result, we note that 
Buda et al.  based  their  conclusions
using image, text, and handwriting examples--
which is much more complex than the data of Table~\ref{tab:datasets} where ``more complex'' can be judged via the number of attributes needed to describe each example:
\bi
\item
Table~\ref{tab:attributes} lists 21 attributes for the SE data;
\item Buda et al. used image data, which are very high-dimensional.
\ei
While the Buda et al. results from row \#2 are disappointing, perhaps they are because the particulars
of Buda et al's method was not based on SE data. Hence, in our own experiments, we tested 
Buda's intuition  (that data sets can be enhanced with oversampling, prior to learning)
even though we do not apply their exact methods. 
This turned out to be a very successful approach since it results in the improvements seen above in 
Figure~\ref{b4after}.

\subsubsection{Optimization with DODGE}
\label{sec:dodge}

One key point to observe from those tables is the size of the optimization space. 

If each numeric range is divided into five bins using {\em (max-min)/5}, and we have 18 options with five choices; i.e. $5^{18}*2^9 \approx 3*10^{12}$ options.

This space is too large to be exhaustively explored. Hence
GHOST   and 
DODGE \cite{agrawal2019dodge} explore this space heuristically.
DODGE's exploration is defined around
 a concept called $\epsilon$-domination.  Proposed by Deb in 2005~\cite{deb2005evaluating}, $\epsilon$-domination states that there is little point in optimizing two options if their performance differs by less than some small value $\epsilon$.

DODGE  exploits   $\epsilon$-domination as follows. $N$ times, repeat: (a)~generate configurations at random, favoring those with the lowest weight (initially, all options have a weight $w=1$); (b)~if one option generates results within $\epsilon$ of some other, then increase these options' weights (thus decreasing the odds that they will be explored again).
This approach can be a very effective way to explore a large space of options.
If we assess a defect predictor  on $d=2$
dimensions (e.g. recall and false alarm), the output space of that performance divides into $(1/\epsilon)^d$ regions.

\subsection{Oversampling, Loss Functions, and Fuzzy Sampling}
\label{sec:loss}
We now discuss the {\em loss functions} we use and our novel oversampling approach
(which led to the superior results
of line \#8 in Table~\ref{tab:ghost-stages}).
In summary, we will apply a {\em fuzzy sampling} method to oversample proportional to the class imbalance ratio, in a way to build robust decision boundaries.

\begin{figure*}[!t]
    \centering
    \begin{subfigure}[t]{0.24\linewidth}
        \centering
        \includegraphics[width=\textwidth]{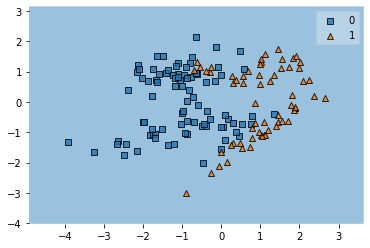}
        \caption{``Vanilla" network}
        \label{fig:wfo:none}
    \end{subfigure}
    \begin{subfigure}[t]{0.24\linewidth}
        \centering
        \includegraphics[width=\linewidth]{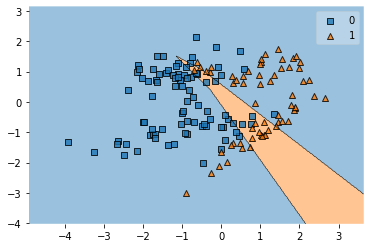}
        \caption{Weighted loss function}
        \label{fig:wfo:weighted}
    \end{subfigure}
    \begin{subfigure}[t]{0.24\linewidth}
        \centering
        \includegraphics[width=\textwidth]{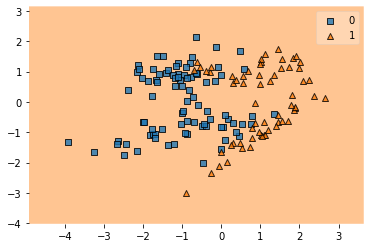}
        \caption{Weighted loss + fuzzy sampling}
        \label{fig:wfo:nosmote}
    \end{subfigure}
    \begin{subfigure}[t]{0.24\linewidth}
        \centering
        \includegraphics[width=\textwidth]{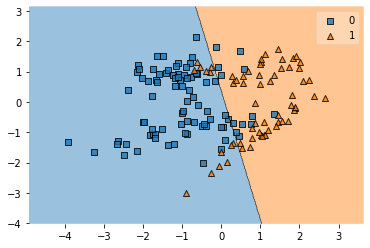}
        \caption{GHOST}
        \label{fig:wfo:final}
    \end{subfigure}
    \caption{An example of how the different components of the GHOST algorithm work. The background color represents the network's predictions, while the points represent the true labels. In each figure is the same dataset artificially generated with some noise (orange is the minority class), and the same feedforward architecture (2 layers, 2 units per layer).}
    \label{fig:wfo-demo} 
\end{figure*}

In neural networks, the loss function $\mathcal{L}: \mathbb{R}^n \times \mathbb{R}^n \mapsto \mathbb{R}$ maps the predictions (given the true labels) to a real number, representing how bad the network is. The \textit{lower} the loss value, the \textit{better}. The loss is used to calculate the gradients that are used to  update the weights of the neural net for the next epoch of learning. 
Loss can be calculated in various ways\footnote{ 
mean squared error,  
binary cross-entropy, 
categorical cross-entropy,
sparse categorical cross-entropy, etc. } but the key point to note here is that, usually, 
{\em the loss function is fixed prior to learning}.

Figure \ref{fig:wfo-demo} illustrates the effects our
various  components. To generate this figure, we created an artificial, imbalanced, noisy classification dataset, where the points represent the true labels, 0 and 1. In the figure, class 0 is the majority, with 67\% of the samples. Based on the results of Montufar et al. \cite{montufar2014number}, we designed a neural network that can perform well on this dataset, with 2 layers, each with 2 units (we use 2 since there are only 2 inputs). 

Figure \ref{fig:wfo:none} shows the result of running this network on the data. Because the data is imbalanced, and the network is trained on a standard loss function (which assumes the goal is accuracy), the network predicts all points as the majority class--we liken this to a student who finds a workaround to get an easy A in a class, while not truly learning anything.

In Figure \ref{fig:wfo:weighted}, we use our weighted loss function to mimic Buda et al.'s recommendation. We see that while this helps, it only caters to a slice of the input space that has a majority of the triangles. 

In Figure \ref{fig:wfo:nosmote}, we also use our novel oversampling technique--now, we have clearly placed far too much emphasis on the minority class, and the learner scores the easy A by exploiting this. 

Our final approach, GHOST, is shown in Figure \ref{fig:wfo:final}. Note how in all the four parts of the figure, the learner (and the data) remains the same; we merely change the \textit{preprocessing}. Clearly, the preprocessing used has a significant impact on the performance of the deep learner, even with its automated feature engineering.

\subsection{Details}

We modify the loss function to adapt
it to the ratio of classes in the data. 

We now elaborate on the weighted loss function.  Consider a loss function $\mathcal{L}(y, \hat{y})$ (where $y$ is a vector representing the target outputs, and $\hat{y}$ is a vector representing the predictions) optimized by a neural network. As is the case with defect prediction, consider the case of binary classification. Let $c_0$ be the minority class, and $C$ be the set of all classes; for binary classification, this is simply the set $\{c_0, c_1\}$. Denote a training example by a subscript (i.e., let $y_i$ denote the target of the $i$th training example), and let $m$ be the number of training examples. Then, the loss function can be rewritten as:

\[
    \begin{aligned}
    \mathcal{L}(y, \hat{y}) &= \sum_{i=1}^m \mathcal{L}(y_i, \hat{y_i}) \nonumber = \sum_{\substack{i=1 \\ y_i = c_0}}^m \mathcal{L}(y_i, \hat{y_i}) + \sum_{\substack{i=1 \\ y_i \neq c_0}}^m \mathcal{L}(y_i, \hat{y_i})
    \end{aligned}
\]
We call the above the \textit{unweighted} loss function. Now, we apply a fairly obvious method of weighting the above as follows: we re-weight the loss function above to:

\begin{equation}\label{eq:w}
\begin{aligned}
    \mathcal{L}^1(y_i, \hat{y_i}) &= \frac{w_i}{n}\sum_{\substack{i=1 \\ y_i = c_0}}^m \mathcal{L}(y_i, \hat{y_i}) + \sum_{\substack{i=1 \\ y_i \neq c_0}}^m \mathcal{L}(y_i, \hat{y_i})
    \end{aligned}
\end{equation}
where $w_i$ is the weight control parameter. For this study, we ran with   various  weights  $w_i  \in \{1,10,100\}$ and found no appreciable difference between these settings. Hence, in the following, we use $w_i=1$. As an aside, we note that this can be extended to multi-class classification.

Clearly, as the minority class shrinks, there is a greater emphasis placed on getting these data points right. This is similar to oversampling by $\frac{w_i}{n}$ because with the unweighted loss function and oversampled data, an individual minority class data point (now represented multiple times) contributes more to the loss function than a majority class data point. However, using this weighted loss function has several advantages (a) we do not need to load additional samples into memory (b) we save computation since we do not need to run iterations over these samples (c) it is flexible: we can use a different weight easily. 

We certainly are not the first to attempt a reweighted loss function: Ryu et al. \cite{ryu2016effective} used weighted Naive Bayes for cross-company defect prediction. More recently, the weighted Naive Bayes was revisited by Arar et al. \cite{arar2017feature} for defect prediction. In the AI literature, Lee et al. \cite{lee2003learning} proposed the use of weighted logistic regression for learning, albeit in the case where some examples are unlabeled (which they consider as negative examples). Lin et al.~\cite{lin2017focal} propose a ``focal loss" function that uses an exponential weighting scheme. However, to the best of our knowledge, we are the first to apply this with various weights in the software engineering domain.

\subsection{Fuzzy sampling}
\label{sec:notation}
Our \textit{fuzzy sampling} triggers on data from the minority classes.
When class frequencies alter greatly between the train and the test
(as seen in Table~\ref{tab:datasets}),
such points are in danger of fall on the wrong side of the border. Therefore, to protect this from happening, we need to push the decision boundary away.

In describing our approach, we use the following notation:
\begin{itemize}
    \item we use $f(\textbf{x}; \theta)$ to denote a neural network $f(\cdot)$;
    \item we denote inputs by $\textbf{x}$;
    \item the neural network function above is \textit{parameterized by} the hyperparameter (and preprocessor) set $\theta$;
    \item we use $\hat{\textbf{y}}$ to denote the predicted values and $\textbf{y}$ to denote the true values;
    \item we use $\hat{\mathcal{L}}$ to denote our modified loss function
    \item we use $m$ to denote the number of training samples;
    \item we use $n$ to represent the fraction of samples belonging to the minority class;
     \item $\boldsymbol{\Delta r}$ is a user-specified parameter. In our experiments, we used values between 0.01 and 0.05 works well.
    \item without loss of generality, we use $c_0$ to denote the minority class, and $c_1$ to denote the majority class;
\end{itemize}

Using the same notation as the weighted loss functions, let $n$ represent the fraction of samples in the minority class $c_0$. Then, iteratively, we add the training samples $(\textbf{x} - i \boldsymbol{\Delta r}, c_0)$ and $(\textbf{x} + i\boldsymbol{\Delta r}, c_0)$ for each sample $(\textbf{x}, c_0)$ in the minority class. At iteration $i$, we add the new samples $\frac{(1 / n)}{2^i}$ times; we iterate till this value is 0. We choose $\boldsymbol \Delta=0.01$, a very small value (to ensure local reasoning). We note here that it might be more beneficial to choose $\Delta r$ based on the statistics of each attribute, but we do not explore that here. Algorithm \ref{alg:wfo} describes the fuzzy sampling sub-routine

\begin{algorithm}[!t]
\footnotesize
    \SetAlgoLined
    \For{each sample $\textbf{x}$ in the minority class}{
        \For{$i \in \{ 0, 1,\ldots\}$ such that $\frac{(1/n)}{2^i} \geq 1$}{
            Add $(\textbf{x} \pm i\Delta r, c_0)$ to the training set $\frac{1/n}{2^i}$ times\;
        }
    }
    \caption{Fuzzy sampling}
    \label{alg:wfo}
\end{algorithm}
\begin{algorithm}[!t]
\footnotesize
    \SetAlgoLined
    \SetKwInOut{KwInput}{Input}
    \SetKwInOut{KwOutput}{Output}
    \KwInput{dataset $D$,\\ performance threshold $\tau = 0.5$, twoSample = \textbf{false}}
    \KwOutput{optimal hyper-parameters $\theta^*$, performance scores $\boldsymbol \phi$}
    Separate $D$ into train and test sets\;
    \If{twoSample} {
        Apply fuzzy sampling to minority class, reversing the class imbalance\;
    }
    Apply fuzzy sampling to the training set\;
    Apply SMOTE to the resulting training set\;
    Choose a set of key hyperparameters and pre-processors\;
     Build a list of options for preprocessing and tuning, and assign every node a weight of 0\;
    Sample at random to create random combinations of preprocessors and hyperparameters (number of layers and units per layer)\;
    \For{$N_1$ random configurations $\theta_1,\ldots,\theta_{N_1}$}{
        \For{$N_0$ epochs}{
            $\hat{\textbf{y}} = f(\textbf{x}; \theta_i)$\;
            $\textbf{w} = \textbf{w} - \nabla_{\textbf{w}} \hat{\mathcal{L}}(\textbf{y}, \hat{\textbf{y}})$\;
        }
        $\phi_i = \text{eval}(f)$\;
    }
    \For{options that result in metrics with a difference $<\epsilon$} {
        Reduce the option weight by 1\;
    }
    \ForAll{other options}{
        Increase the weight by 1\;
    }
    \For{$N_2$ evaluations}{
        Choose the options $\theta_i$ with the highest weight\;
        \For{$N_0$ epochs}{
            $\hat{\textbf{y}} = f(\textbf{x}; \theta_i)$\;
            $\textbf{w} = \textbf{w} - \nabla_{\textbf{w}} \hat{\mathcal{L}}(\textbf{y}, \hat{\textbf{y}}$)\;
        }
    }
    $\theta^* = \text{argmax}_{\theta} \text{eval}(f)$, $\boldsymbol \phi = \max \text{eval}(f)$\;
    \If{$\boldsymbol \phi < \tau$ \textbf{and} twoSample = \textbf{false}}{
        Run GHOST with twoSample = \textbf{true}\;
    }
    \KwRet{$\theta^*$}
    \caption{GHOST}
    \label{alg:ghost}
\end{algorithm}

and Algorithm \ref{alg:ghost} describes our overall approach:
\bi
% \item
% Note line 8 is where we would apply hyper-parameter optimization (which would, in turn, runs lines 8 and and above many
% times find good parameters).
\item
Lines 11 to 14 are the deep learner (and here, we use the standard deep learner described above).
Based on the results of Theorem 5 of Montufar et al. \cite{montufar2014number}, we use
a 2-layer network with 20 units per layer.
\item
Lines 17 to 22 shows DODGE's tabu search  (discussed in \S\ref{tuneit}. The weights of
similar results get reduced such that, in subsequent reasoning (see line 24), we do not waste our time sampling from that region.
\item
Lines 23 to 30 re-runs DL  to find good configurations.
\item Line 31, 32 say that if we are failing more
than half the time $\tau < 0.5$ then try again,
this time doubling up of the fuzzy sampling.
\ei

\noindent
As pointed out above, Figure \ref{fig:wfo-demo}  contains several components (DODGE, SMOTE, etc).
It is hence reasonable to ask if any of those components are superfluous.
To check the value
of each component, we ran the ablations study shown in Table~\ref{tab:ghost-stages}. Note that the best results came from using all our components, and the largest increase in performance was realized by adding in fuzzy sampling.

\BLUE
%\respto{2a2.1}
The method of oversampling twice (when \textit{twoSample = \textbf{true}}) leads to a significantly larger dataset. Specifically, if the original training set contains $m$ minority class samples, and $mk$ majority class samples, i.e., the class imbalance ratio is $k$, then it can be shown that the total number of points when \textit{twoSample} is \textbf{true} has an upper bound of $7mk+3m$, which is over 3 times as large as the original dataset ($\frac{7mk+3m}{m+km}= 3+\frac{4k}{k+1}$. This is important because this increase is also reflected in the runtime of the algorithm. In our testing, we noticed that with \textit{twoSample = true}, although GHOST is significantly better, it runs 2-3x slower than if it were false, consistent with this result. Therefore, in Algorithm \ref{alg:ghost}, we first run with \textit{twoSample = false}, and, in line 31, check if our performance is sub-optimal; if so, we run it again with \textit{twoSample = true}.
\BLACK

\subsection{Software engineering and GHOST}
\label{sec:se-and-ghost}

In this section, we discuss the applicability of GHOST to software engineering and non-SE tasks, such as image processing.

Because GHOST is fundamentally based on DODGE, we argue that the applicability of GHOST is directly related to the applicability of DODGE. Specifically, because DODGE uses a tabu search mechanism, it is limited to lower-dimensional hyper-parameter spaces. We note that both in the DODGE paper and in our experiments, this condition is met--we tune the preprocessor, the number of layers, and the number of units in each layer, forming a 3-dimensional space. However, in complex domains such as image recognition, this is not possible due to the sophisticated nature of the architectures required to achieve state-of-the-art results. This result has been empirically proven recently by Agrawal et al.~\cite{agrawal2019ai}, who show that DODGE performs poorly in standard AI datasets, such as those found in the UCI repository. For this reason, we discuss our approach only in the context of software engineering (and, in particular, defect prediction), with small hyper-parameter search space dimensionality.

\BLUE
\subsection{When \textit{not} to use GHOST?}
\label{sec:pitfalls}
%\respto{3a3.1}
With any approach, it is equally important to understand its limitations. In this section, we list two potential pitfalls of using GHOST, and how to fix them in practice.

\begin{enumerate}[(a)]
    \item \textbf{Trespassing:} Note that GHOST relies heavily on the use of our novel fuzzy sampling approach, which concentrically adds points around minority samples to push the decision boundary away. In doing so, these newly added samples can inadvertently cross a majority sample, effectively adding more noise to the dataset, and confusing the learner further. If this \textit{``trespassing"} occurs to many of the minority class samples, it could deteriorate the performance of the learner. This is easily fixable, by using smaller values of $\Delta r$.
    \item \textbf{Balanced datasets:} In balanced datasets, fuzzy sampling only adds additional computational overhead in an attempt to balance classes that do not require it. In such cases, we recommend that practitioners simply use hyper-parameter tuning to yield good results.
\end{enumerate}
\BLACK

\section{Experiments}
\label{sec:method}

The ablation study of Table~\ref{tab:ghost-stages} was a  heuristic method to quickly confirm
that no part of our proposed oversampling method is superfluous. That done, we can now explore 
several research questions, in depth.

{\bf RQ1: How well does standard DL perform versus the prior state-of-the-art?} 

We ask
this question since we are critical of
 SE researchers that merely explore DL without baselining this new method against prior non-DL methods.
As we shall see, in such a comparison, standard DL performs  poorly.
 
{\bf RQ2: How can we fix  DL for defect prediction?} 

The above discussion around Figure~\ref{fig:wfo-demo}
conjectured  that  (a)~widely varying class ratios between train and tests sets introduce
instabilities into the boundary between classes; and (b)~this problem can be alleviated
by oversampling around ``at risk'' members of the minority classes.  To check this conjecture,
we apply the system designed above, which resulted in large improvements in DL performance.

{\bf RQ3: How scalable is  this fixed version of DL?} 

Any method that tries tuning deep learning (as we do)
adds a significant computational burden to an algorithm that is already very slow.
Hence it is appropriate to check the scalability of our methods. We show that our methods
are scalable since a 400\% growth in the size of the data results in a less than 180\%
growth in the runtimes.

\subsection{Experimental Rig}

Our experiments  compared different defect predictors. For  the prior non-DL 
state-of-the-art  we used the  methods from DODGE \cite{agrawal2019dodge} that appeared
in a 2019 TSE paper. 

For DL, we use a standard deep learner (from \S\ref{sec:deeplearning}) with 2 layers and 20 units per layer, trained for 30 epochs (these settings were taken from Theorem 5 in Montufar et al. \cite{montufar2014number}).

For a state-of-the-art  deep learner, we used
the 
Wang et al. \cite{wang2018deep} method  from a 2018 TSE paper (also described in \S\ref{sec:deeplearning}).
We    use Wang et al. since (a)~that paper made a convincing case that this approach represented state-of-the-art results for defect prediction for
 features extracted from source code and  (b) a deep learner is used to extract features, rather than being used as the learner, in contrast to our approach; this forms a valuable comparison.

We also explored two variants of defect prediction:
\bi
\item 
{\em Within-project defect prediction (WPDP).} Here, models
learned from earlier releases of some project  predict properties
of latter releases of that project.
\item
{\em Cross-project defect prediction (CPDP).} Here, models learned
from project1 were then applied to project2. 
\ei

When comparing against DODGE,  we split the data into train and test sets, as shown in Table \ref{tab:datasets}.  GHOST used the training sets to find good DL settings, which were then applied to the test suite. 

As stated above, while we oversample the training data, we {\em never} modify the test data (since that would we mean are not testing on the kinds of data that might be encountered  in the field).

Because of the stochastic nature of deep learning (caused by random weight initialization), a statistically valid study of its merits should consider the distribution of its performance. For this reason, we run GHOST 20 times for each dataset, for each metric. All the results reported (in Table \ref{tab:results} and Table \ref{tab:tan}) are median values of these 20 runs.

\subsection{Data}
For this study, we use the same data used in DODGE's prior study~\cite{agrawal2019dodge} on defect prediction: see Table \ref{tab:datasets}. These are all open-source Java projects from the PROMISE repository \cite{Sayyad-Shirabad+Menzies:2005}.  Each dataset has 20 static attributes for each software project, described in Table \ref{tab:attributes}.

% In our subsequent discussion, the
% following feature of  Table \ref{tab:datasets} will become very important. In this table,  the  class balance    between defective and non-defective modules is highly variable between our train and test sets:
% \bi
% \item For example,   in velocity  and jedit, the percent of defect modules decreases more than half between train and test sets;
% \item The reverse pattern is seen in the train/test sets of other data sets.
% For example,  the percentage of defect modules
% increases by (approx) 50\% (for poi and synapse) and by over 300\% (for log4j,xalan, xerces). 
% \ei
%  This feature will  motivated the design of the GHOST method, described later in this paper.

\subsection{Performance Metrics}\label{sec:perfs}
For this study, we use   performance metrics widely used in 
the defect prediction literature.
Specifically, we evaluate the performance of our learners using three metrics: Area Under the ROC Curve (AUC) with the false positive rate on the x-axis and true positive rate on the y-axis, recall, false alarm rate, and popt20. We use these measures to compare against the baseline of Agrawal et al.~\cite{agrawal2019dodge}.
\bi
\item
Recall measures the number of positive examples that were correctly classified by a learner.
% :
\item
% \[
%     \text{recall} = \mathit{TP}\;/\;(\mathit{TP}+\mathit{FN})
% \]
% where TP is the number of true positives and FN is the number of false negatives.
The false alarm rate (which we denote by \textit{pf}) is the fraction of negative samples a classifier incorrectly classifies as a positive example.
%Hence, we will use {\em pf}, defined as follows:
% \[
%      \text{pf} =  \mathit{FP}\;/\;(\mathit{FP}+\mathit{TN}) %= %\frac{\text{TN}}{\text{FP}+\text{TN}}
% \]
% where FP is the number of false positives and TN is the number of true negatives. 
\item
The next metric, popt20 comments on the inspection effort required after a defect predictor is run. To calculate popt20, we list the defective modules before the non-defective modules, sorting both groups in ascending order of lines of code. Charting the percentage of defects that would be recalled if we traverse the code sorted in this manner in the y-axis, we report the value at the 20\% point. 
\item
Finally, we report the area under the receiver operating characteristics (ROC) curve (AUC), with false positive rate on the x-axis and true positive rate on the y-axis.
\ei

A key benefit of using multiple metrics, aside from a more complete view of the model's performance, is that it allows us to more concretely test whether deep learning works for all situations. Since some metrics are more important for certain stakeholders (e.g., popt20 comments on the effort needed after the defect predictor are run; others may prefer a combination of high recalls and low false alarm rates, i.e., d2h), this exploration allows us to point out the cases where deep learning may not be the most suitable option.

\subsection{Statistics}
\label{sec:stats}

The statistical methods used in this paper were selected according to the particulars of our experiments.

For example, in  the {\em within-project experiments},  we are using learners that employ stochastic search. When testing such algorithms, it is standard~\cite{arcuri11} to repeat  those runs 20 times with different random number
seeds. Such  experiments generate a distribution of  20 results per learner per data set.
 For those experiments, we use
 the {\em distribution  statistics} of \S\ref{sec:dist}
 to compare the efficacy
 of different learners.
 
 For the cross-project experiments, we are comparing our results against the methods of Wang et al. \cite{wang2016automatically},
 As mentioned in \S\ref{sec:algdesign},  we do not have access to their implementations but we do have access to the train and test sets they use. For
 these experiments, we must compare our results to the single performance points mentioned in the  Wang et al. paper \cite{wang2016automatically}. 
 For those experiments, we use the {\em point  statistics}
 of \S\ref{sec:point}
 to compare the efficacy
 of different learners.

\subsubsection{Distribution statistics}\label{sec:dist}

Distribution statistics \cite{arcuri13parameterto, ghotra2015revisiting} are used to distinguish two distributions of data. In our experimental setup, we run GHOST and DODGE 20 times each, and therefore have a distribution of results for each optimizer. This allows us to use distribution statistical methods to compare results.

Our comparison method of choice is the Scott-Knott test, which was endorsed at TSE'13 \cite{mittas2012ranking} and ICSE'15 \cite{ghotra2015revisiting}. The Scott-Knott test is a recursive bi-clustering algorithm that terminates when the difference between the two split groups is insignificant. Scott-Knott searches for split points that maximize the expected value of the difference between the means of the two resulting groups. Specifically, if a group $l$ is split into groups $m$ and $n$, Scott-Knott searches for the split point that maximizes

\[
    \mathbb{E}[\Delta] = \frac{|m|}{|l|}\left( \mathbb{E}[m] - \mathbb{E}[l] \right)^2 + \frac{|n|}{|l|}\left( \mathbb{E}[n] - \mathbb{E}[l] \right)^2
\]

where $|m|$ represents the size of the group $m$.

The result of the Scott-Knott test is \textit{ranks} assigned to each result set; higher the rank, better the result. Scott-Knott ranks two results the same if the difference between the distributions is insignificant.

\subsubsection{Point statistics}\label{sec:point}
% ,

% Point statistics are used when point samples of two results are available. When comparing against the baselines of Wang et al. \cite{wang2016automatically}, because we do not have 20 runs of their algorithm (because of the proprietary nature of their code and extracted features--see Section \ref{sec:dlse}), we choose to use point statistics to check if one result is better than another.

% For all our results, we report the median over 20 runs. When comparing between different

For point statistics, we have access to various performance points (e.g. recall) across multiple data sets.
To determine if one point is better than another, we have to define a delta $\Delta$
below which we declare two points are the same.

To that end, we
use recommendations from
Rosenthal~\cite{rosenthal1994parametric}.
%(as of June 2020,  this paper has 2713  citations in Google Scholar).  
Rosenthal comments that for point statistics, 
parametric methods have more statistical power (to distinguish groups) than nonparametric methods. Further, within the parametric methods, there are two families of methods: 
  those that use the
  ``$r$''  Pearson product moment correlation;
  and
  those that use some ``$d$'' normalized difference between the mean.

After making those theoretical points Rosenthal goes on to remark that  neither method is intrinsically better than another.
Using Rosenthal's advice, we apply the most straightforward method endorsed by that research.   Specifically, we compare  treatment performance differences using Cohen's delta, which is computed as
$\Delta = 0.3*\sigma$
where $\sigma$ is the standard deviation of all the values seen across all the data sets.
When two methods are different by less than $\Delta$, we say that they {\em  perform equivalently}. 
Otherwise,  we say that one method {\em out-performs} the other if its performance is larger than
$\Delta$.

\section{Experimental Results}
\label{sec:results}

The rest of this paper uses the above methods to explore the research questions listed in
the introduction.

\subsection{How well does standard DL perform versus the prior state-of-the art? ({\bf RQ1})}
\label{sec:rq1}

We are critical of
 SE researchers that merely explore DL without baselining this new method against prior non-DL
 state-of-the-art.   Accordingly,
 before we do anything else, we make that comparison.

Table \ref{tab:results} compares DL versus the prior state-of-the-art (DODGE) versus
our preferred new method (GHOST). 
The wins, ties, and losses from that table are summarized on
Table~\ref{tab:ghost_dodge}. Note that these wins, tie, losses
were computed using the statistical methods of
\S\ref{sec:stats}.

As can be seen in the 40 experiments of   Table \ref{tab:results}, DL only  performs  better than or the same as
DODGE in  10/40 experiments (see all the entries marked with ``*''). 
 A repeated pattern is that when DL wins on false alarms, it usually does so at the expense
 of low recalls.  Also,   when it does achieve a high recall (e.g. for velocity),
 this usually comes at the cost 
 of  high false alarms.
Hence we say:

\newcommand{\lightgray}[1]{\cellcolor{blue!7}\textcolor{black}{#1}}
\newcommand{\gray}[1]{\cellcolor{blue!20}\textcolor{black}{#1}}

\begin{table}[!t]
\caption{RQ1 and RQ2 results.
Cells show medians of 20 runs.
Cells with an  "*" show the few
cases where DL
worked better than DODGE.
 \colorbox{blue!20}{Dark blue} shows  top rank (note: for pf, {\em less}
 is {\em better}), \colorbox{blue!7}{light blue} shows rank two;  white shows  lowest rank (worst performance). 
Rankings were calculated via
\S\ref{sec:dist}. The train and test versions used here are the same as the DODGE paper (see Table \ref{tab:datasets}).
}\scriptsize
\label{tab:results}
\centering {\scriptsize
\begin{tabular}{ll|llll}
\toprule
                                         &       & AUC  & popt20 & recall & pf   \\ \midrule
\multicolumn{1}{l}{\multirow{3}{*}{ivy}} & DL &  0.58 & 0.15 & 0.33 & \gray{0.27} * \\
\multicolumn{1}{c}{}                     & DODGE & \gray{0.71} & \lightgray{0.25}   & \lightgray{0.85}   & \lightgray{0.36} \\
\multicolumn{1}{c}{}                     & GHOST & \lightgray{0.69}  & \gray{0.31}   & \gray{0.95}    & \lightgray{0.38} \\
\midrule \multirow{3}{*}{camel}          & DL & 0.55 & \lightgray{0.18} & 0.10 & \gray{0.04} *\\       
                                         & DODGE & \lightgray{0.58} & \gray{0.54}   & \lightgray{0.63}   & \lightgray{0.36} \\
                                         & GHOST & \gray{0.62} & \gray{0.54}   & \gray{0.65}   & 0.41 \\
\midrule \multirow{3}{*}{jedit}          & DL & 0.61 & 0.10 & 0.46 & \gray{0.16} *\\         
                                         & DODGE & \lightgray{0.63} & \lightgray{0.39}   & \lightgray{0.64}   & \gray{0.25} \\
                                         & GHOST & \gray{0.68} & \gray{0.41}   & \gray{0.91}   & \lightgray{0.40} \\
\midrule \multirow{3}{*}{log4j}          & DL & 0.55 & \lightgray{0.31} & 0.33 & \gray{0.13} *\\        
                                         & DODGE & \lightgray{0.61} & \gray{0.99}   & \lightgray{0.54}   & \lightgray{0.22} \\
                                         & GHOST & \gray{0.66} & \gray{0.99}   & \gray{0.75}   & \gray{0.19} \\
\midrule \multirow{3}{*}{velocity}       & DL & 0.50 & 0.57 & \gray{0.90} *& 0.87 \\         
                                         & DODGE & \lightgray{0.61} & \gray{0.64}   & 0.76   & \gray{0.47} \\
                                         & GHOST & \gray{0.68} & \lightgray{0.64}   & \lightgray{0.82}    & \gray{0.47}  \\
\midrule \multirow{3}{*}{synapse}        & DL & 0.54 & \lightgray{0.23} & 0.20 & \gray{0.05} *\\         
                                         & DODGE & \lightgray{0.65} & \gray{0.48}   & \gray{0.65}   & \lightgray{0.23} \\
                                         & GHOST & \gray{0.67} & \gray{0.48}   & \lightgray{0.63}   & 0.33 \\
\midrule \multirow{3}{*}{lucene}         & DL & 0.58 & \lightgray{0.51} & \gray{0.69} *& 0.52 \\         
                                         & DODGE & \gray{0.61} & \gray{0.80}   & \gray{0.67}    & \lightgray{0.36}  \\
                                         & GHOST & \lightgray{0.59} & \gray{0.80}   & \gray{0.7}   & \gray{0.34}  \\
\midrule \multirow{3}{*}{xalan}          & DL & 0.55 & \lightgray{0.24} & 0.21 & \gray{0.09} *\\        
                                         & DODGE & \lightgray{0.71} & \gray{1.0}    & \lightgray{0.71}   & \lightgray{0.14}  \\
                                         & GHOST & \gray{0.75} & \gray{1.0}    & \gray{0.76}   & 0.27 \\
\midrule \multirow{3}{*}{xerces}         & DL & 0.52 & \lightgray{0.28} & 0 & \gray{0.04} *\\         
                                         & DODGE & \lightgray{0.59} & \gray{0.93}   & \lightgray{0.54}   & \lightgray{0.15}  \\
                                         & GHOST & \gray{0.62} & \gray{0.94}   & \gray{0.57}   & 0.39 \\
\midrule \multirow{3}{*}{poi}            & DL & \lightgray{0.61} & 0.36 & \lightgray{0.45} & \gray{0.18} *\\         
                                         & DODGE & \gray{0.72} & \lightgray{0.66}   & \gray{0.78}   & \lightgray{0.22} \\
                                         & GHOST & \gray{0.73} & \gray{0.74}   & \gray{0.78}   & 0.38  \\
                                         \bottomrule
\end{tabular} }
\end{table}
\begin{table}[!t]
    \centering
    \caption{RQ2 results (our DL method vs the prior non-DL state-of-the-art).
    This table summarizes  Table~\ref{tab:results}.
    The first column indicates the number of wins, ties,  and losses  for each metric (these are defined using the $\Delta$ measure of Section \ref{sec:perfs} and the directions for ``better'' defined in Section \ref{sec:stats}.  Note that for popt20 and pf, there are multiple ties because both DODGE and GHOST achieve the highest possible score.}
  
  \scriptsize
   \begin{tabular}{rllll|r} 
   \toprule
    & AUC & Recall & popt20 & pf & Total \\
    \midrule
     \textbf{win} & 7 & 6 & 3 & 2 & 18 \\
    \textbf{tie} & 1 & 3 & 6 & 1 & 11 \\
     \textbf{loss} & 2 & 1 & 1 & 7 & 11 \\ \midrule 
     \textbf{win + tie} & 8 & 9 & 9 & 3 & 29 \\
     \bottomrule
    \end{tabular}
    \ 
    \label{tab:ghost_dodge}
\end{table}

\begin{blockquote}
    \noindent
    \BLUE
    %\respto{2a1.1}
    For defect prediction,  standard deep learning is sub-optimal.
    \BLACK
\end{blockquote}

 Note that these {\bf RQ1} results merely show that standard deep learning performs badly.
 Those results do not explain {\em why} that is so. For this task,
 we must move on to {\bf RQ2}.

\subsection{  How to Fix Deep Learning? {\bf (RQ2)}}
\label{sec:rq2}

The above discussion around Figure~\ref{fig:wfo-demo}
showed  that  (a)~widely varying class ratios between train and test sets introduce
instabilities into the boundary between classes; and (b)~this problem can be alleviated
by oversampling around ``at risk'' members of the minority classes.  To check this conjecture,
we apply the system designed above.

\begin{table}[!t]
\centering
\caption{RQ2 results. Comparison of GHOST (optimized for recall) with the results of Wang et al. \cite{wang2016automatically}, using deep learning-generated ``Semantic" features to train a Naive Bayes model on within-project defect prediction. Bold indicates better results, as determined by the point statistics of \S\ref{sec:point} (and here, $0.3 \sigma = 4.91$).}
\label{tab:wpdp} {\scriptsize
\begin{tabular}{l|ll|lll|lll}
\toprule
Dataset & Train & Test & \multicolumn{3}{c}{Wang et al.} & \multicolumn{3}{c}{GHOST}    \\
\midrule
                         &                        &                                          & P    & R    & F1             & P    & R    & F1                \\
                         \midrule
\multirow{2}{*}{synapse} & 1                      & 1.1                                   & 46   & 66.7 & 54.4           & 74.5 & 98.5  & \cellcolor{blue!10}\textbf{84.6*}              \\
                         & 1.1                    & 1.2                                  & 57.3 & 59.3 & 58.3           & 66.4 & 100  & \cellcolor{blue!10}\textbf{79.8*}              \\
                         \midrule
\multirow{2}{*}{jEdit}   & 3.2                    & 4                                      & 46.7 & 74.7 & \cellcolor{blue!10}\textbf{57.4}           & 42.3 & 100  & \cellcolor{blue!10}\textbf{55.9}              \\
                         & 4                      & 4.1                                   & 54.4 & 70.9 & 61.5           & 74.7 & 100  & \cellcolor{blue!10}\textbf{85.5*}              \\ \midrule
log4j                    & 1                      & 1.1                                   & 67.5 & 73   & \cellcolor{blue!10}\textbf{70.1}           & 55.6 & 100  & \cellcolor{blue!10}\textbf{65.9}              \\ \midrule
ivy                      & 1.4                    & 2                                     & 21.7 & 90   & 35             & 88.6 & 100  & \cellcolor{blue!10}\textbf{83.4*}              \\ \midrule
\multirow{2}{*}{lucene}  & 2                      & 2.2                                   & 75.9 & 56.9 & 65.1           & 61.3 & 100  & \cellcolor{blue!10}\textbf{74.5}              \\
                         & 2.2                    & 2.4                             & 66.5 & 92.1 & \cellcolor{blue!10}\textbf{77.3}           & 60.9 & 100  & \cellcolor{blue!10}\textbf{75.3}              \\ \midrule
\multirow{2}{*}{camel}   & 1.2                    & 1.4                           & 96   & 66.4 & 78.5           & 83.4 & 100  & \cellcolor{blue!10}\textbf{90.9*}              \\
                         & 1.4                    & 1.6                                  & 26.3 & 64.9 & \cellcolor{blue!10}\textbf{37.4}           & 26.7 & 100  & \cellcolor{blue!10}\textbf{38.2}              \\ \midrule
xalan                    & 2.4                    & 2.5                                  & 65   & 54.8 & 59.5           & 62.7 & 100  & \cellcolor{blue!10}\textbf{66}                \\ \midrule
xerces                   & 1.2                    & 1.3                               & 40.3 & 42   & 41.1           & 84.8 & 100  & \cellcolor{blue!10}\textbf{91.8*}                \\ \midrule
\multirow{2}{*}{poi}     & 1.5                    & 2.5                                & 76.1 & 55.2 & 64             & 72.2 & 100  & \cellcolor{blue!10}\textbf{83.2}              \\
                         & 2.5                    & 3                                     & 81.6 & 79   & \cellcolor{blue!10}\textbf{80.3}           & 70.2 & 100  & \cellcolor{blue!10}\textbf{79.7}              \\ \midrule
\multirow{2}{*}{ant}     & 1.5                    & 1.6                                  & 88   & 95.1 & \cellcolor{blue!10}\textbf{91.4}           & 81.3 & 93.8 & 87.1*              \\
                         & 1.6                    & 1.7                                  & 98.8 & 90.1 & \cellcolor{blue!10}\textbf{94.2}           & 52.6 & 99.4 & 57.3              \\
 \bottomrule
\end{tabular} }
\end{table}

\begin{table}[!t]
 \caption{RQ2 results. Summary of the 
    Table~\ref{tab:wpdp} GHOST vs  Wang et al. results.
    Comparison of GHOST with the results of Wang et al. \cite{wang2016automatically} on cross-project defect prediction. All results shown are F-1 scores. GHOST results are medians over 20 runs. Bold and \colorbox{blue!10}{dark blue} indicates better (in case of tie, both are in bold), as determined by point statistics
     of \S\ref{sec:point} (and here, 
    $0.3 \sigma = 4.41$).}
    \label{tab:tan}
    \centering
     {\scriptsize
    \begin{tabular}{ ll|ll } 
    \toprule
	\textbf{Source} & \textbf{Target} & \textbf{Wang et al.} & \textbf{GHOST} \\
	\midrule
	camel-1.4 & jedit-4.1 & \cellcolor{blue!10}\textbf{61.5}  & 47.3 \\ 
	lucene-2.2 & log4j-1.1 & \cellcolor{blue!10}\textbf{61.8}  & 51.7 \\ 
	synapse-1.2 & ivy-2.0 & \cellcolor{blue!10}\textbf{82.4}  & 30.3 \\
	camel-1.4 & ant-1.6 & \cellcolor{blue!10}\textbf{97.9}  & 84.9* \\
	xalan-2.5 & xerces-1.3 & \cellcolor{blue!10}\textbf{38.6} & \cellcolor{blue!10}\textbf{35.7} \\ 
	jedit-4.1 & log4j-1.1 & \cellcolor{blue!10}\textbf{64.5} & \cellcolor{blue!10}\textbf{64.1} \\
	log4j-1.1 & jedit-4.1 & 50.3 & \cellcolor{blue!10}\textbf{85.5*} \\
	xerces-1.3 & ivy-2.0 & 45.3 & \cellcolor{blue!10}\textbf{94.0*} \\
	jedit-4.1 & camel-1.4 & 69.3  & \cellcolor{blue!10}\textbf{90.9*} \\ 
	ivy-2.0 & xerces-1.3 & 42.6 & \cellcolor{blue!10}\textbf{91.8*} \\ 
	lucene-2.2 & xalan-2.5 & 55  & \cellcolor{blue!10}\textbf{65.4} \\ 
	xerces-1.3 & xalan-2.5 & 57.2 & \cellcolor{blue!10}\textbf{66} \\ 
	xalan-2.5 & lucene-2.2 & 59.4 & \cellcolor{blue!10}\textbf{73.8} \\ 
	log4j-1.1 & lucene-2.2 & 69.2 & \cellcolor{blue!10}\textbf{74.6} \\ 

	ivy-2.0 & synapse-1.2 & 43.3 & \cellcolor{blue!10}\textbf{60.9} \\ 
	poi-3.0 & synapse-1.2 & 51.4 & \cellcolor{blue!10}\textbf{58.1} \\ 
	synapse-1.2 & poi-3.0 & 66.1 & \cellcolor{blue!10}\textbf{82.7} \\ 
	ant-1.6 & camel-1.4 & 31.6 & \cellcolor{blue!10}\textbf{36} \\ 
	ant-1.6 & poi-3.0 & 61.9 & \cellcolor{blue!10}\textbf{77.8} \\ 
	poi-3.0 & ant-1.6 & 47.8 & \cellcolor{blue!10}\textbf{62.5} \\ 
	\bottomrule 
\end{tabular}
}
    
\end{table}

\begin{table}
\centering
    \caption{RQ2 results. Statistical  of the 
    Table~\ref{tab:wpdp} and Table \ref{tab:tan} GHOST vs  Wang et al. results~\cite{wang2016automatically}.
    Generated using the methods of \S\ref{sec:stats}.}
    \label{tab:tan-summary} { \scriptsize
    \begin{tabular}{l|p{0.9cm}p{0.9cm}|l}
    \toprule 
         & \textbf{WPDP} & \textbf{CPDP} & \textbf{Total} \\
         & (Tbl. \ref{tab:wpdp}) & (Tbl. \ref{tab:tan}) & \\
         \midrule 
       win  & 9 & 14 & \textbf{23} \\
       tie  & 5 & 2 & \textbf{7} \\
       loss & 2 & 4 & \textbf{6} \\
       \midrule 
       \textbf{win + tie} & 14 & 16 & 30 \\
       \bottomrule
    \end{tabular} }
\end{table}

Returning to
Table \ref{tab:results}, GHOST can be seen
to have more
 \colorbox{blue!20}{dark blue}
 cells than anything else (i.e. statistically, it is ranked number one most often).
Table~\ref{tab:ghost_dodge} 
%and Table~\ref{tab:ghost_dl} 
summarizes those results for comparison with DODGE:
%\bi
%\item
   18 times, GHOST defeats DODGE (see top-right, Table \ref{tab:ghost_dodge});
%\item
%  29 times, GHOST defeats  DL (see top-right, Table \ref{tab:ghost_dl});\item
%  11 times, GHOST was as good as DODGE (see mid-right,  Table \ref{tab:ghost_dodge})
%\ei
Tables~\ref{tab:results} and \ref{tab:ghost_dodge}
%\ref{tab:ghost_dl} 
display distribution results where algorithms  were run multiple times using different random number seeds.
Tables   \ref{tab:wpdp}
and \ref{tab:tan},
on the other hand,
 display
point distribution results where our algorithms are compared to the single set of
performance points
reported in prior work.

In Table \ref{tab:wpdp} and Table \ref{tab:tan}, we show the comparison of GHOST with the results of Wang et al. In these tables, in cases where we used \textit{twoSample} to improve our results, we denote such results with an asterisk. As shown in the summary (Table \ref{tab:tan-summary}), we perform as good or better in 14/16 datasets for within-project defect prediction. In cross-project defect prediction, we also note that we win 14/20 times and tie 2/20 times; this suggests that GHOST may be able to generalize well across different projects. 

Table \ref{tab:tan-summary} summarizes the comparison with the results of Wang et al. \cite{wang2016automatically}.

These results recommend GHOST over DODGE and DBN. Also, they deprecate the use of off-the-shelf standard DL for defect prediction (since GHOST clearly is preferred to standard DL). We attribute the super performance of GHOST to its weighted loss function.

The exceptions to the above pattern are the  GHOST vs DODGE recall results, which we will discuss in the next section.
Apart from that, we say that:

\begin{blockquote}
    \noindent
    For most evaluation goals, our modified version of deep learning (GHOST) performs better than the prior state-of-the-art.
\end{blockquote}

\subsection{Scalability  (RQ3)}
\label{sec:rq3}

\begin{wraptable}{r}{0.4\linewidth}
    \caption{Median training time over 20 repeats.}
    \label{tab:runtimes}
    \scriptsize
    \begin{tabular}{rll}
        \toprule
          & Training   & GHOST   \\
           &   (secs) &  (secs) \\
        \midrule
        ivy & 0.78 & 9m 17s \\
        lucene & 0.81 & 10m 26s \\
        poi & 1.09 & 13m 44s \\
        synapse & 0.81 & 9m 48s \\
        velocity & 0.88 & 10m 20s \\
        camel & 1.41 & 18m 10s \\
        jEdit & 1.11 & 14m 32s\\
        log4j & 0.73 & 8m 29s\\
        xalan & 1.61 & 20m 16s\\
        xerces & 1.08 & 11m 56s \\
        \bottomrule \\
    \end{tabular}
\end{wraptable}

Any method that tries tuning deep learning (as we do)
adds a significant computational burden to an algorithm that is already very slow.
Hence it is appropriate to check the scalability of our methods. 

To address this issue, we report the median training time over 20 runs. The measured time is CPU time, when trained on a 4-core Intel Core i5 CPU. These are summarized in Table \ref{tab:runtimes}. Clearly, our models are very fast (less than 2 seconds to train on a CPU).

Table \ref{tab:runtimes} also shows the runtimes for running GHOST on different datasets. Because GHOST runs 20 times to find the median value of a metric, the runtimes we report are for all 20 runs. However, we do not divide this time by 20 as we feel it is scientifically important to run a stochastic experiment multiple times and report statistical results.

\begin{figure}[!t]
    \centering
    \includegraphics[width=.9\linewidth]{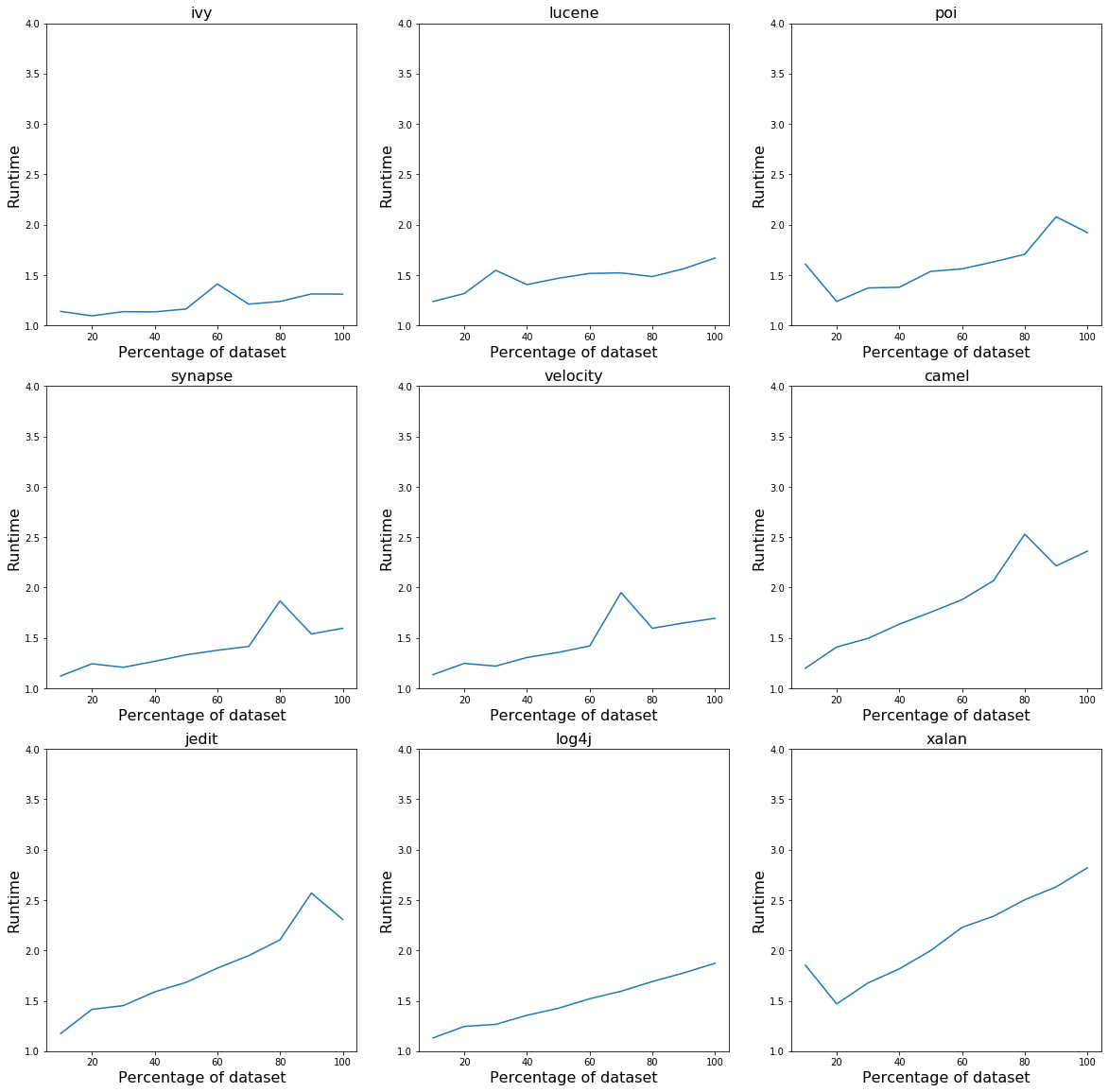}
    \caption{RQ3 results: scalability results for 9 datasets. The x-axis is the perentage of the dataset trained on; the y-axis is the runtime in seconds on a CPU.}
    \label{fig:scalability}
\end{figure}

Figure \ref{fig:scalability} shows our scalability results, which we obtain by measuring the training time for different sizes of the datasets.
We observe a general trend across all datasets that deep learning scales well with the size of the problem. More specifically,  GHOST's runtimes grow sublinearly. For example, in the xalan results, 
 a 400\% increase in data (from a fifth to all the data) leads to a runtime increase of   only    $2.7/1.5 = 180\%$.

Hence we say: 
\begin{blockquote}
    \noindent
    Tuning deep learners is both practical and tractable for defect prediction.
\end{blockquote}

\section{Threats to Validity}
\label{sec:threats}

\textbf{Sampling bias:} As with any other data mining paper, it is important to discuss sampling bias. We claim that this is mitigated by testing on 10 SE projects over multiple versions, and demonstrating our results across all of them. Nevertheless, in future work, it would be useful to explore more data. 

\textbf{Learner bias:} Our learner bias here corresponds to the choice of architectures we used in our deep learners. As discussed above, we chose the architectures based on our reading of ``standard DL'' from the literature. That said, different DL architectures could lead to different results.

\textbf{Evaluation bias:} We compared our methods using a range of   metrics: AUC, recall, false alarm rate, and popt20. We used these  varied evaluation metrics to demonstrate more clearly the benefits of not using learners off-the-shelf, and tuning the loss function optimized. If other performance metrics are used, then other results might be obtained.

\textbf{Order bias:} This refers to a bias in the order in which data elements appear in the training and testing sets. We purposely choose these so that the testing set is from a newer version than the data used in the training set, following the natural order of software releases. Therefore, we argue that such a bias is needed to mimic real-world usage of such learners.

\textbf{External validity:} We tune major hyperparameters using DODGE, removing external biases from the approach. Our baseline results are based on the results of Montufar et al. ~\cite{montufar2014number}, which has been evaluated by the deep learning community.

\section{Conclusion}
\label{sec:conclusion}
There is a discrepancy in the DL 
literature between those
who argue that DL automates
feature engineering~\cite{zeiler2014visualizing,panda2016unsupervised,nair2010rectified,suk2014hierarchical,yamashita2018convolutional}  and the
results of Buda et al.~\cite{buda2018systematic},
which advocate an oversampling
pre-processor. To the best of our knowledge,
prior to this paper,
that discrepancy has not been previously
explored using SE data.

We find that:

\begin{blockquote}
    \noindent
    Oversampling is effective and necessary prior to applying deep learning for defect prediction.
\end{blockquote}

Through extensive experimentation motivated by both prior work \cite{montufar2014number} and visualization (see Figure \ref{fig:wfo-demo}), we made a case that class imbalance is a problem for deep learners, and the need for oversampling as a preprocessing step \textit{prior} to relying on the automated feature engineering offered by deep learning. Our literature review showed that this was lacking in software analytics; our paper is the first to present a cogent, thorough case for (a) comparing against strong, non-DL baselines (b) hyper-parameter tuning for maximum performance of the learners used (c) oversampling to alleviate the class imbalance problem.

Our results showed that while standard random oversampling does not suffice, a novel adaptive version \textit{combined} with the recommendations of prior work \cite{montufar2014number,agrawal2019dodge,Chawla02} yields state-of-the-art results much faster. The approach we proposed, GHOST, is scalable and works across a wide variety of defect prediction datasets. Further, through ablation studies, we demonstrated the necessity of each component of the algorithm.

Therefore, GHOST can be summarized as a new method that combines deep learning with 
with two useful extensions:  (a) weighted loss functions (see \S\ref{sec:loss}) (b) a novel fuzzy sampling approach (see \S\ref{sec:notation}) (c) SMOTE \cite{Chawla02}.
 We demonstrated the efficacy of our approach on 10 defect prediction datasets, using four metrics, as well as on within-project and cross-project defect prediction data as studied by Wang et al. \cite{wang2016automatically}. We compared our results against multiple baselines: (a) a standard
 deep learner and  (b) a prior state-of-the-art result in defect prediction using non-deep learning methods, and (c) a prior state-of-the-art result using deep learning to extract features from code.  
 We presented scalability and runtime tests to demonstrate that deep learners can be trained quickly.
We showed that the best results come from  learners properly tuned for the dataset and the goal metric (using a weighted loss function). 

We note that one could apply the weighted loss functions to classical learners, such as support vector machines (SVMs), which rely on an optimization setup, which has certainly been tried before \cite{ryu2016effective, arar2017feature, lee2003learning,lin2017focal}. However, the ablation studies showed that the most significant improvements to our performance came from our novel fuzzy sampling approach. Moreover, the motivation for fuzzy sampling (i.e., to prevent the boundary from getting too close to the training data) also applies to learners such as decision trees.

 We take  care to stress that our results relate to defect prediction. As to other areas of software analytics, that is a matter for future search. That said,
our results suggest that, for future work, the following research agenda could be insightful:
 \begin{enumerate}
\item
 Divided analytics into various domains. 
 \item
 Then, for each domain:
 \begin{enumerate}\item
 Determine the prior non-DL state-of-the-art; \item Compare DL with that state-of-the-art; 
 \item Seek ways to better adapt DL to that domain.
 \end{enumerate}\end{enumerate}

\section*{Acknowledgements}
This work was partially funded by 
a research grant from the National Science Foundation (CCF \#1703487).

\balance
\bibliographystyle{IEEEtran}
{\small \bibliography{cite} }

\begin{minipage}{.45\textwidth}
% biography
\begin{IEEEbiography}[{\includegraphics[width=1.05in,clip,keepaspectratio]{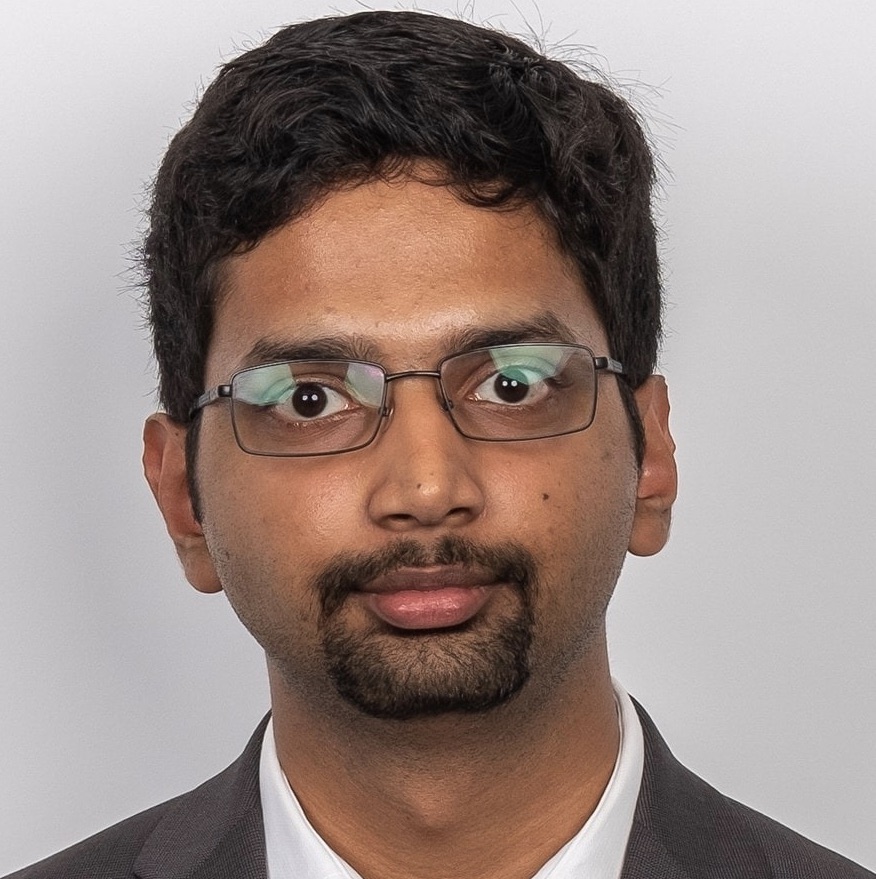}}]{Rahul Yedida} is a PhD student in Computer Science at NC State University. His research interests include automated software testing and machine learning for software engineering. For more information, please visit \url{https://ryedida.me}.
\end{IEEEbiography}

\begin{IEEEbiography}[{\includegraphics[width=1.05in,clip,keepaspectratio]{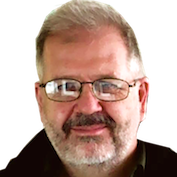}}]{Tim Menzies} (IEEE Fellow, Ph.D. UNSW, 1995)
is a Professor in computer science  at NC State University, USA,  
where he teaches software engineering,
automated software engineering,
and programming languages.
His research interests include software engineering (SE), data mining, artificial intelligence, and search-based SE, open access science. 
For more information,  please visit \url{http://menzies.us}.
\end{IEEEbiography}
\end{minipage}

\end{document}